\documentclass[twocolumn,amssymb,amsmath,pre,showpacs]{revtex4}

\usepackage{rotating,epsfig,dcolumn} 
\usepackage{graphicx,bm}

\begin{document}

\title{$1/f$ noise in the Two-Body Random Ensemble}

\author{A. Rela\~no$^a$, R. A. Molina$^b$, and 
J. Retamosa$^a$}
\affiliation{
(a) Departamento de F\'{\i}sica At\'omica, Molecular y Nuclear, 
 Universidad Complutense de Madrid 28010 Madrid, Spain.\\
(b) CEA/DSM, Service de Physique de l'Etat Condens\'e,
 91191 Gif-sur-Yvette, France.}

\date{\today}

\begin{abstract}
We show that the spectral fluctuations of the Two-Body Random Ensemble
(TBRE) exhibit $1/f$ noise.  This result supports a recent conjecture
stating that chaotic quantum systems are characterized by $1/f$ noise
in their energy level fluctuations. After suitable individual
averaging, we also study the distribution of the exponent $\alpha$ in
the $1/f^{\alpha}$ noise for the individual members of the
ensemble. Almost all the exponents lie inside a narrow interval around
$\alpha=1$ suggesting that also individual members exhibit 1/f noise,
provided they are individually unfolded
\end{abstract}


\maketitle

Our understanding of quantum chaos has greatly advanced during the
last two decades. The pioneering work of Berry and Tabor
\cite{Berry:77}, and Bohigas, Giannoni and Schmit \cite{Bohigas:84}
showed that there exists a close relationship between the energy level
fluctuation properties of a quantum system and the large time scale
behavior of its classical analogue. In their seminal paper, Bohigas
{\it et al.}  conjectured that the fluctuation properties of generic
quantum systems which in the classical limit are fully chaotic
coincide with those of random matrix theory (RMT). This conjecture is
strongly supported by experimental data, many numerical calculations,
and analytical work based on semiclassical arguments. Later the
interest in these studies was renewed with the discovery that the
spectral statistics of quantum disordered systems is also well
described by RMT. A review of later developments can be found in
references \cite{Guhr:98,Stockmann:99}.

Thus, RMT plays a fundamental role in quantum chaos studies, though it
was originally introduced by Wigner to describe the statistical
properties of high-lying energy levels of quantum systems
\cite{Wigner:51}. To describe spectral fluctuations, RMT assumes that
physical Hamiltonians can be substituted by ``reasonable'' random
ensembles of Hamiltonian matrices, and introduces convenient
statistics of their level spectra. RMT should provide the ensemble
average of these statistics. Usually, these statistics
are the nearest neighbor spacing distribution \cite{Mehta:91}
introduced to analyze the short range correlations, and the Dyson
$\Delta_3$ \cite{Mehta:91} statistic that allows to study long range
correlations.

Recently a new approach, based on traditional methods of
time series, has been proposed to analyze spectral fluctuations
\cite{Relano:02}.  In this first work we showed that the classical random
matrix ensembles (CRME) exhibit $1/f$ noise in the fluctuations of the
excitation energy.  We also presented evidences that this is actually
a universal property of quantum chaotic systems. The purpose of this
brief report is to study whether the $1/f$ noise present in the
spectral statistics of the CRME is also present in the Two-Body Random
Ensemble (TBRE). As we shall see below, this question is very
pertinent if we want to apply this statistic to the study of the
spectral fluctuations of real many-body systems.
 
The CRME, usually called Gaussian and Circular ensembles, were chosen
due to their invariant properties under certain symmetry
transformations \cite{Mehta:91}. For example, the Gaussian Orthogonal
Ensemble (GOE) is invariant under orthogonal transformations and is
applicable to systems invariant under time-reversal symmetry. However,
the GOE represents systems with N-body interactions while normal
systems in nature are supposed to be very well described by effective
two-body interactions in the mean-field basis. TBRE was introduced to
tackle this problem, and in this sense is more appropriated to study
atomic nuclei, quantum dots and other mesoscopic systems. {\em This
ensemble is constructed from a GOE in the 2-particle Hilbert space and
then propagating it to the N-particle Hilbert space by using the
direct product structure of this type of spaces}.  (for that reason
this kind of ensembles are also called Embedded GOE (EGOE))
\cite{French:70,Bohigas:71a}. Given the single particle states
$|v_i>,\;\; i\in{1,2,\cdots,M}$, the two-body Hamiltonian is written
as

\begin{equation}
H = \sum_{v_i<v_j,v_k<v_l} <v_k v_l|H|v_i v_j>
a^{\dagger}_{v_l}a^{\dagger}_{v_k}a_{v_i}a_{v_j},
\label{H2}
\end{equation}

\noindent
where $a^{\dagger}_{v_i}$ ($a_{v_i}$) creates (destroys) a fermion in
the state $|v_i>$. The two body matrix elements $<v_k v_l|H|v_i v_j>$
are properly antisymmetryzed and are taken to be independent
Gaussian random variables with 

\begin{eqnarray}
\overline{<v_k v_l|H|v_i v_j>} & =& 0, \nonumber\\
\overline{\left|<v_k v_l|H|v_i v_j>\right|^2} & =& \sigma^2(1+\delta^c_{(kl),(ij)}).
\label{TBME}
\end{eqnarray}
In this equation $\overline{\;\cdot\;}$ denotes ensemble average, $\sigma$
is a constant and $\delta^c$ is the Kronecker delta. Then, the
Hamiltonian matrix in the N-particle space is defined in terms of
these two-body matrix elements via the direct product structure. The
only non-zero N-particle matrix elements are of three types

\begin{eqnarray}
<v_1 v_2 \cdots v_N |H|v_1 v_2 \cdots v_N > & =& \sum_{v_i<v_j\le v_N} <v_i v_j|H|v_i v_j> \nonumber \\
<v_p v_2 \cdots v_N |H|v_1 v_2 \cdots v_N > & =& \sum_{v_i= v_2}^{v_N} <v_p v_i|H|v_1 v_i> \nonumber \\
<v_p v_q \cdots v_N |H|v_1 v_2 \cdots v_N > & =& <v_p v_q|H|v_1 v_2>,
\label{MBME}
\end{eqnarray}

\noindent
or those obtained by permuting the single particle states. All other
matrix elements are zero.

Very few analytic results are known for TBRE, contrary to the
classical random matrix ensembles. A very important result is that the
level density of the TBRE is Gaussian in the dilute limit, which
corresponds to $(N,M)\longrightarrow \infty$, $N/M\longrightarrow 0$
\cite{French:70,Bohigas:71a}, instead of the semicircular law for the
GOE \cite{Mehta:91}. To perform a numerical analysis of the TBRE
spectral statistics an important difficulty must be overcome: the TBRE
is not ergodic \cite{Bohigas:71b,French:73,Brody:81}. In the present
context ergodicity means that the statistical properties of individual
ensemble members (and hence those of the physical Hamiltonian) should
always coincide with the ensemble average. In order to transform TBRE
into an ergodic ensemble the spectrum of each member must be unfolded
(see the unfolding description below) individually. In this way, GOE
statistics is recovered. For a recent review of TBRE and more
generally EGOE see reference \cite{Kota:01}

In order to establish whether $1/f$ noise is also present in the
spectral fluctuations of TBRE, we have studied four ensembles with
different matrix sizes. We have treated $N=6$ ``spinless'' fermions in
$M=11$, $12$, $13$ and $14$ degenerated states leading to Hilbert
space dimensions $D=462$, $924$, $1716$ and $3003$ respectively.  The
TBRE matrices were constructed using eqs. (\ref{H2}), (\ref{TBME}) and
(\ref{MBME}) with $\sigma=1$. There is no relevant energy scale in the
model and the only parameter is the dimension of the Hilbert space. We
have diagonalized $200$ matrices in each case to obtain the ensemble
average.

For each Hamiltonian matrix the level density $\rho(E)$
can be separated into a smooth part $\overline{\rho}(E)$, that defines
the main trend of the level density, and a fluctuating part
$\widetilde{\rho}(E)$. It is well known that level fluctuations
amplitudes are modulated by $\overline{\rho}(E)$; therefore, to
compare the statistical properties of different systems or different
parts of the same spectrum, the main trend defined by
$\overline{\rho}(E)$ must be removed. This procedure, called
unfolding, consists in mapping the level energies $E_i$ into new
dimensionless levels $\epsilon_i$,

\begin{equation}
E_i \longrightarrow \epsilon_i = \overline{\Sigma}(E_i),\; i=1,2,\cdots,D,
\end{equation}
where $\overline{\Sigma}(E)$ is an smooth approximation to the actual
step function $\Sigma(E)$ that gives the true number of energy levels
from the ground state energy $E_0$ and up to energy $E$. This function
is given by
\begin{equation}
\Sigma(E) = \int_{E_0}^E \rho(\eta) d\eta.
\end{equation}
We have already commented the important analytical result of French
and Wong who showed that in the TBRE the mean level density
$\overline{\rho}(E)$ goes to Gaussian form in the dilute limit.
However, for the dimensions of the matrices used in this work, the
corrections to the Gaussian behavior are very important and different
for each matrix. Since the use of an accurate unfolding
procedure is essential to avoid misleading results for the long range
spectral correlations \cite{Gomez:02}, we have selected another
method.  Recently, the problems related to the unfolding procedure in
the TBRE have been discussed deeply in
Refs. \cite{Flores:00,Jackson:01}. After some tests we have chosen polynomials
up to grade 5 to fit the accumulated level
density $\Sigma(E)$; higher grades produce spurious long-range
correlations.  Finally, we have thrown $5\%$ of the eigenvalues in the
two spectrum edges.

 In the approach of Ref. \cite{Relano:02}, the analogy of the energy
spectrum with a time series is established in terms of the $\delta_q$
statistic. Using the unfolded energies it is defined as
\cite{Mehta:91}.

\begin{equation}
\delta_q=\sum_{i=1}^q\left(s_i-\left<s\right>\right) = \epsilon_{q+1}-\epsilon_1-q.
\end{equation}
where $s_i$ are the next-neighbor level spacings,$s_i=\epsilon_{i+1} -
\epsilon_i$, with spectral average value $\left<s\right>=1$.  Note
that $\delta_q$ represents the deviation of the excitation energy of
the $(q+1)$-th unfolded level from its mean value. Moreover, it is
closely related to the level density fluctuations. Indeed, we can
write
\begin{equation}
\delta_q = \overline{\Sigma}(E_{q+1})-\Sigma(E_{q+1}) = -\widetilde{\Sigma}(E_{q+1}), 
\label{def_delta_2}
\end{equation}
if we appropriately shift the ground state energy; thus, it represents
the accumulated level density fluctuations at $E = E_{q+1}$.

We will profit of the formal similarity of the $\delta_q$ function
with a time series to analyze its properties with numerical
techniques, normally used in the domain of complex systems. The most
simple procedure is to study the scaling properties of its power
spectrum $S(k)$. The latter is defined in terms of the discrete
Fourier transform

\begin{equation}
\widehat{\delta}_k=\frac{1}{d}\sum_{q=1}^{d-1} \delta_q \exp \left( \frac{2 \pi i k q}
{d}\right)
\end{equation}

\noindent
in the usual way as 

\begin{equation}
S(k)=\left|\hat{\delta}_k\right|^2,
\end{equation}

\noindent
where $d\le D$ is the total number of unfolded levels considered.  In
the present work $d \simeq 0.9 D$ We will say that the
spectral fluctuations of a Hamiltonian ensemble exhibit $1/f$ noise if
the ensemble averaged power spectrum of $\delta_q$ follows a power law
of type

\begin{equation}
\overline{S(k)} \propto \frac{1}{k^{\alpha}},
\end{equation}

\noindent
with $\alpha\simeq 1$. For a single Hamiltonian is not clear whether
we must impose that the bare power spectrum or some kind of average
follows the previous power law. We shall explore three different
possibilities below.

\begin{figure}[ht]
\begin{center}
\psfig{file=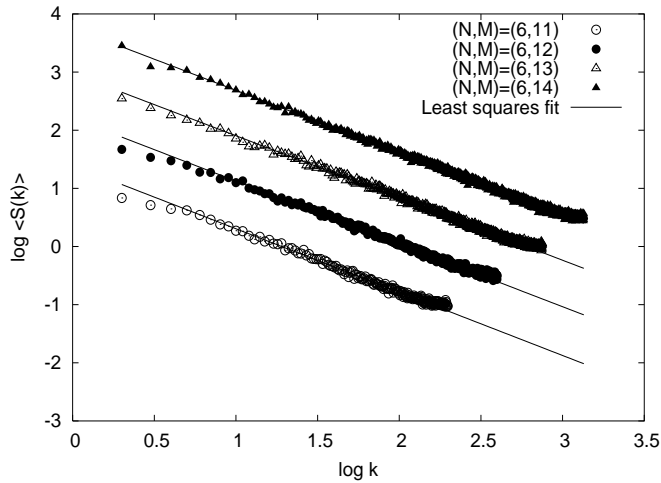,height=9cm,width=6.5cm,angle=-90}
\caption{Ensemble averaged power spectra of four TBRE with different
dimensions. The best $1/f^{\alpha}$ fit is also shown.  The curves
have been displaced vertically to avoid overlapping between them.}
\label{Fig1}
\end{center}
\end{figure}

The results obtained for the ensemble averaged power spectra are shown
in Fig. \ref{Fig1} using a log-log scale. It clearly seen that the
calculated points spread along straight lines. The line slopes, i.e.,
the power spectrum exponents are obtained by means of a least-squares
fit and their values are $\alpha=1.09 \pm 0.04$, $\alpha=1.08 \pm
0.01$, $\alpha=1.07 \pm 0.01$ and $\alpha=1.07 \pm 0.01$ for
$(N,M)=(6,11)$, $(6,12)$, $(6,13)$ and $(6,14)$, respectively. The
exponents are very close to one, confirming that there is $1/f$ noise
in the TBRE. Thus, we obtain a new and powerful check
of the conjecture that links the spectral statistics of the TBRE with
that of the GOE.

\begin{figure}[t]
\begin{center}
\psfig{file=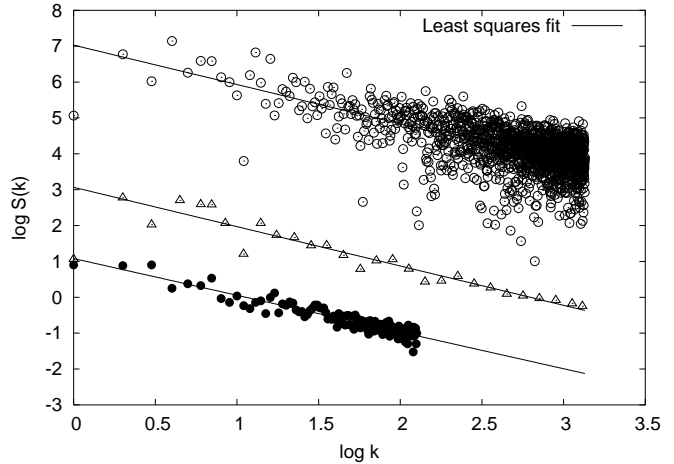,angle=-90,width=9cm}
\caption{Three examples of power spectra for an individual Hamiltonian
selected from the $(N,M)=(6,14)$ TBRE. The curves have also been
displaced vertically to avoid overlapping.  Open circles represent the
raw power spectrum, open triangles the average values of the power
spectrum obtained by binning the log (frequency) scale, and filled
circles the running averaged power spectrum. The solid line represents
the best fit in each example. The power spectrum exponents are
$\alpha= 1.10\pm 0.07$, $\alpha=1.10 \pm 0.06$, and $\alpha= 1.02\pm
0.05$ respectively.}
\label{Fig2}
\end{center}
\end{figure}

Moreover, in order to study to what extent the
$\delta_q$ statistic is also meaningful for individual spectra, we
have randomly selected a member pertaining to the TBRE ensemble with
$(N,M)=(6,14)$.  The upper plot of Fig \ref{Fig2} shows the power
spectrum of the $\delta_q$ function for this member. Although this
result suggests the existence of a power law, the calculated points
are widely spread around the mean behavior, and therefore other
different curves can be used to fit the data points. Performing a
least square fit to a straight line we obtain a value $\alpha = 1.10
\pm 0.07$, but the error seems unreliable. Following B. Mandelbrot,
the problem arises because of the double logarithmic plot: {\em
spectral components must never be plotted raw, only after suitable
averaging} \cite{Mandelbrot:99}.  One of the best procedures to
perform this average consists in dividing the high frequency portion
of the logarithmic frequency axis into equal bins and averaging the
power spectrum components in each bin. The result of this procedure is
shown in the second plot of fig. \ref{Fig2}: the averaged data points
are no more widely spread, but all of them fall near the mean
behavior. If we perform a least square fit to this new set of data
points we obtain $\alpha=1.10 \pm 0.06$; therefore, the fuzzy behavior
is confirmed to be related just to the double logarithmic plot. An
alternative averaging procedure is to calculate a {\em running or
spectral average}, $\left< S(k)\right>$. Since in this case the
dimension is large enough, we can divide the whole spectrum in $10$
different sets of $256$ consecutive levels. Then, the $\delta_q$ power
spectrum is calculated for each level set and in order to reduce
fluctuations and clarify the main trend a running average is performed
using these sets. The bottom plot of Fig. \ref{Fig2} displays the
result of this calculation using open circles.  A least squares fit
leads to the following exponent $\alpha= 1.02\pm 0.05$, which is very
similar to that obtained by means of an ensemble average. This result
is compatible with the ergodic properties of TBRE once the spectra are
individually unfolded. Note that the different lengths of the three
power spectra shown in this figure are due to the fact that in the
first and the second cases we are actually using the whole sequence,
but in the third one we use sequences of $256$ consecutive levels.

\begin{figure}[t]
\begin{center}
\psfig{file=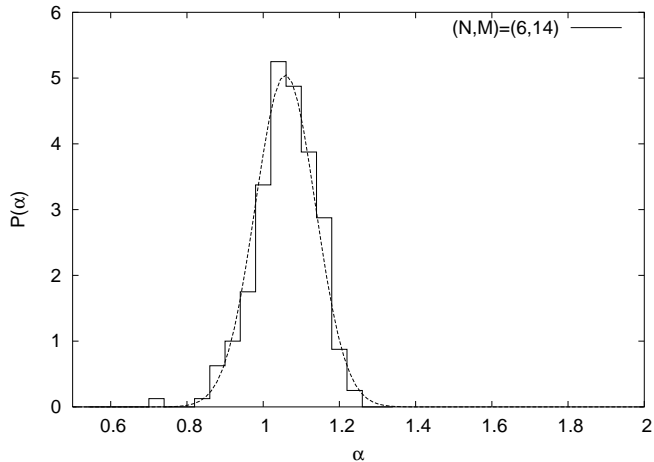,angle=-90,width=9cm}
\caption{Histogram of the distribution of $\alpha$ for the
$(N,M)=(6,14)$ TBRE.}
\label{Fig3}
\end{center}
\end{figure}

To make this discussion more quantitative, we have calculated the
exponent $\alpha$ for the 200 matrices of the ensemble
with $(N,M)=(6,14)$ using the binning method previously described.
The average value is $<\alpha> = 1.06$ and the width of
the distribution is $\sigma_{\alpha}=0.08$. Figure \ref{Fig3} shows an
histogram of the distribution together with a Gaussian defined by
the previous parameters that seems to fit the data very well.
Although this result has been obtained for a particular sample of
a particular ensemble $((N,M)=(6,14))$, it suggests that individual
members also are characterized by $1/f$ noise.

We have confirmed that the spectral fluctuations of the TBRE exhibit
$1/f$ noise. This behavior supports the previously stated conjecture
that chaotic quantum systems are characterized by $1/f$ noise in their
energy level fluctuations. We have also shown that individual members
have $1/f$ noise in their excitation energy fluctuations provided they
are individually unfolded and the power spectrum of the $\delta_q$
function is appropriately averaged. Actually, the
distribution of the $\alpha$ exponent in the $1/f^{\alpha}$ law is a
Gaussian centered near $\alpha=1$ with a quite small
width. Therefore, the spectral fluctuations of atomic nuclei, quantum
dots and mesoscopic systems can be studied by means of the scaling
properties of the power spectrum of the $\delta_q$ function.  The
advantages of this new statistic are perfectly used in some recent
works about the nuclear masses \cite{Hirsch:1,Hirsch:2}, where the
$1/f$ noise of different series of fluctuations in the nuclear masses
along the nuclear chart was explored. Depending of the definition of
the fluctuations, different exponents in the power law were
found. $1/f$ noise was shown to be a powerful tool to investigate
spectral correlations in this kind of experimental data.

\end{document}